\documentclass{ws-procs9x6}

\def\apj{{ApJ }}                 % Astrophysical Journal
\def\aap{{A\&A }}                % Astronomy and Astrophysics
\def\memras{{MmRAS }}            % Memoirs of the RAS
\def\nat{{Nature }}              % Nature
\def\eqalign#1{\,\vcenter{\openup.7ex\mathsurround=0pt
\ialign{\strut\hfil$\displaystyle{##}$&$\displaystyle{{}##}$\hfil
 \crcr#1\crcr}}\,}

\newcommand{\be}{\begin{eqnarray}}
\newcommand{\ee}{\end{eqnarray}}

\begin{document}

\title{Signatures of  middle aged, nearby pulsars in the cosmic ray lepton spectrum?}

\author{I. B\"usching}

\address{Theoretische Physik IV, Ruhr-Universit\"at Bochum,\\
44801 Bochum, Germany\\
$^*$E-mail: ib@tp4.rub.de}

\author{Okker C. deJager}

\address{Unit for Space Physics, North-west University,\\ 
        Potchefstroom 2520, South Africa\\
        $^*$E-mail:Okkie.DeJager@nwu.ac.za}

%\begin{abstract}
%\end{abstract}
\bodymatter
\section{Introduction}
 Recent data reported by the PAMELA \cite{Adriani09} and ATIC \cite{atic08} experiments, as well as cosmic ray lepton results from FERMI \cite{ferminew} and H.E.S.S. \cite{hess2} collaborations sparked  a series of papers explaining these results either by contributions of electron positron pairs to the local interstellar cosmic ray (CR) spectrum by  dark matter (DM) or pulsars. Focusing here on pulsars, we argue that at the present, our knowledge about particle  acceleration at pulsars  as well as of the local Galactic CR propagation is still limited, i.e. the recent results for  CR electrons and positrons constrain pulsar and propagation models. 
We will thus not present another attempt to explain the data by contributions of pulsars to the local CR lepton flux but rather to highlight the caveats in doing so.
\section{Pulsars as Particle Accelerators}
Modelling the acceleration of particles at pulsars and subsequent injection of these particles into the interstellar medium, one may look at
three scenarios:
young pulsars (after the breakup of their pulsar wind nebulae)\cite{buesching08}, mature pulsars %\cite{chi96},
\cite{zhang01} and millisecond pulsars\cite{buesching08c}. 
\subsection{Energy Budget}
Before discussing the acceleration of particles by pulsars, it is instructive to look for the available energy budget of a single pulsar. The energy reservoir of a pulsars is its rotational energy $1/2\Omega^2$, where $I$ is the moment of inertia. %It draws its power the spin down. % of the pulsar. 
The spin down power $L_{\rm SD}=I\Omega\dot\Omega$  of a pulsar is given by\cite{reesgunn74}
\be
L_{\rm SD}(t) &=& L_{\rm sd,0}\left(1+{t}/{\tau_0}\right)^{-\frac{n+1}{n-1}},
\label{eq:Lsd}
\ee
with the characteristic decay time $\tau_0= P_0/((n-1)\dot{P}_0)$ and $n = 3$ 
for a magnetic dipole field of the pulsar.
 
With $P_0$, $\dot{P}_0$ being the period and its derivative at pulsar birth 
and 
assuming the magnetic field of the pulsar does not decay, the spin down power at birth  can be written as
\begin{equation}
L_{\rm sd,0} =\left| -\frac{4\pi^2I\dot{P}P}{P_0^4}\right|.
\label{eq:Lsd0a}
\end{equation}
It is easily seen from Eq.~\ref{eq:Lsd0a} that, given a small enough birth period, one can get 
an arbitrary large energy reservoir. 10\,ms can be seen as a reasonable lower limit for $P_0$, while initial periods in the range $\approx$50--150\,ms are not uncommon \cite{faucherkaspi06}. It is estimated that up to 40\% of the pulsars may be born with periods in the range 100--$500$\,ms \cite{vranesevic04}.

\subsection{Source Spectra}
Several models have been put forward  to describe the  acceleration of CR leptons by pulsars (see e.g \cite{profumo} and references therein).

The simplest approach is to assume that some fraction $f_{\rm part}$ of the spin-down power $L_{\rm SD}(t)$ is transferred to particles, which follow a power-law spectrum with index $a$ and  exponential cut-off.
The acceleration of leptons at  $\gamma$-ray pulsars has been modeled by a number of authors. Whereas  \cite{harding_ramaty} assume that the positron spectrum follows the $\gamma$-ray spectrum, it is assumed by
\cite{zhang01} that mature pulsars with ages larger than 100\,kyr inject their mono-energetic wind into the interstellar medium.  

From Eq.~\ref{eq:Lsd} it is clear that the available spin-down power (and thus the available energy that can be transferred to particles) is largest when the pulsar is still young. In fact, integrating  Eq.\ref{eq:Lsd} over time for $n=3$, it becomes clear that  at $t=\tau_0$ the pulsar dissipated  half of its available rotational energy. One can  therefore expect that the majority of particle acceleration at a given pulsar takes place for $t>\tau_0$. What is indeed observed are  nebula of relativistic particles around many young pulsars, i.e. pulsar wind nebulae (PWN). There is also evidence, that these highly relativistic particles are confined in the PWN \cite{dejager09a}.   

Assuming the particles from the pulsar are reaccelerated at the  pulsar wind shock and then contained in the PWN until its breakup, 
\cite{buesching08} model the particle spectrum injected into the interstellar medium by the two middle aged, nearby pulsars B0656+14 and Geminga.
We generalize this model by assuming a broken power law particle spectrum at the pulsar shock \cite{venterdejager06}
\be
Q^\prime(E,t)&= \left\{\eqalign { k^\prime(t) \left({E}/{E_B}\right)^{-2} \,\,\,\,\,\qquad {\rm for\,\, } E >E_B
\cr
 k^\prime(t)\left({E}/{E_B}\right)^{-1} \qquad {\rm for\,\, } E <E_B \cr} \right.
\ee
keeping $E_B$ a free parameter. 
The spectrum at the shock  
is normalised by the available energy deposited into  particles 
\begin{equation}
\int_{E_{\rm min}}^{E_{\rm max}} Q^\prime(E,t)EdE  =  f_{\rm part}L_{\rm SD}(t).
\label{eq:energy}
\end{equation}
The fraction of the spin down power deposited in particles
\begin{equation}
f_{\rm part}=\eta/{(1+\sigma)}
\end{equation} 
can be expressed in terms of the  magnetisation parameter
\be
\sigma(t)&=&0.003\left({t}/{1\,{\rm kyr}}\right)^{3/2}.
\ee
With  $\epsilon = r_L/r_{\rm shock} = 0.001\dots0.1$, the maximum particle energy at shock is  \cite{venterdejager06}
\begin{equation}
E_{\rm max}=\epsilon e \kappa \sqrt{{\sigma}/({\sigma+1})} \sqrt{{L_{\rm SD}}/{c}}.
\end{equation}
$\kappa=3$ is the compression ratio at the shock and $e$ the elementary charge. 

The evolution of the particle spectrum in the PWN is governed by synchrotron losses
\be
\frac{\partial Q(E,t)}{\partial t}-  Q^\prime(E,t)=\frac{\partial}{\partial E}\left(B_{\rm PWN}(t)^2E^2Q(E,t)\right)
\label{pwnevo}
\ee
where the mean  PWN magnetic field is assumed to decay with time
\be
B_{\rm PWN}(t)&=&{1200}/{\left(1+t/{\rm kyr}\right)^2}\,\,\, [{\rm \mu G}]. 
\ee
Under these assumptions, the particle spectrum in the PWN at time of breakup $T$  is:
\be
Q(E,T)&=&\int_0^{T}Q^{\prime}(E_0,t_0)E_0^2E^{-2}\Theta\left(E_0-E_{\rm min}\right)\Theta\left(E_{\rm max}-E_0\right)dt_0,
\label{geminga:spec}
\ee
with
\be
 E_0&=&{E}/\left({E\int_t^{t_0}B_{\rm PWN}(t^{\prime})^2dt^{\prime}+1}\right)
\nonumber
\ee
\section{Propagation of Cosmic Ray leptons}
The propagation of CR leptons in the Galaxy is dominated by diffusion and energy losses due to synchrotron radiation in the Galactic magnetic field.  The differential density $N$ satisfies the  equation
%\begin{equation}
\be
\frac{\partial N}{\partial t}-S = \nabla \cdot\left(k \nabla N\right)
- \frac{\partial}{\partial E}\left( b_0E^2 N\right),
\label{ana:pgl}
\ee
%\end{equation}
where $b_0$ is determined by the Galactic magnetic field.

In the case of a pulsar, the 
source function  is given by
%\begin{equation}
\be
S=Q(E,t)\delta\left(\vec{r}-\vec{r}_{\rm source}\right), 
\ee
%\end{equation}
where $Q(E,t)$ is given by a model for particle acceleration at pulsars. 
Above $\approx 4\,$GV, the diffusion coefficient has an energy  dependence of the form 
%\begin{equation}
\be
k=k_0\left({E}/{1~{\rm GeV}}\right)^{\alpha},
\label{efiff:k}
\ee
%\end{equation}
where $\alpha=0.3\dots0.7$ and  $k_0$ are determined by fitting observed secondary to primary and radioactive secondary data. Depending on the model, $k_0$ is in the range 0.006\,kpc$^2$Myr$^{-1}$--0.2\,kpc$^2$Myr$^{-1}$ (e.g. \cite{maurin02}).

Although obtaining  $k_0$ in this way is a well established procedure, we want to remark the following. First, fitting  secondary to primary (e.g B/C) data only yields a correlation between $k_0$ and the halo height $H$ \cite{maurin01}. This degeneracy can be broken by looking at  
radioactive to stable secondary ratios (e.g. $^{10}$Be/Be). In doing so, one compares  life times due to  radioactive decay with the times scale of escape, the latter depends on the diffusion perpendicular to the Galactic disk. Thus the fitting of the observed chemical composition of CR by a propagation model measures primarily the propagation perpendicular to the Galactic disk, which not necessarily has to be the same as that in the Galactic plane. 
Second, to the best of our knowledge, all studies determining  $k_0$ by fitting the observed chemical composition so far use 1D or 2D models. Nevertheless, it has been shown \cite{buesching05}, that the primary component of locally observed CR flux may change at least by a factor of two depending on the local SN history, given these objects are indeed the main sources of Galactic CR. The flux of the secondary component on the other hand shows little or no  variation. Thus, the observed secondary to primary ratios will depend on the local SN history. It may be necessary to  focus on tertiary to secondary ratios when working with leaky box, 1D or 2D propagation models. 
%
%In case of an infinite diffusion volume, which is a good approximation for CR leptons with energies above several 100\,GeV Eq.~\ref{ana:pgl} can be solved analytically by the 
%Green's function formalism (see e.g.  \cite{berezinskii90}, \cite{atoyan95}).
\section{Modelling the Contribution of Pulsars to the local Cosmic Ray Lepton Spectrum} 
If the source function of electrons and positrons accelerated at a given pulsar and also the distance to this object is known, one can calculate the contribution of these particles to the locally observed CR spectrum.
As discussed above, the magnitude of the diffusion coefficient in the Galactic plane is not well determined. Also, the spectrum of the particles injected into the interstellar medium strongly depends on the model used.

Focusing on particle acceleration at young pulsars as, according to Eq.~\ref{eq:Lsd}, these objects promise the largest output of high energy leptons, we follow the model outlined in the previous section.
%We focus now on 
%the model of  B\"usching, deJager and Venter, as it  considers particle acceleration at young pulsars, and thus, according to Eq.~\ref{eq:Lsd}, promises the larges output of leptons.   
%The  parameters of this model span a wide parameter space. 
According to pulsar physics, we have to allow for a large range in parameter space.
The pulsar birth period can be in the range  $\approx$ 10\,ms to 500\,ms,  $E_B$, is most probably in the range 10--1000\,GeV, and  the PWN  life time $T$ smaller 100\,kyr. 
Within this parameter space in addition to the uncertainties in $k_0$, it is possible to
produce a wide range of particle spectra at earth
which can fit the data. In fact, looking at Geminga or B0656+14 individually,  the available CR lepton data imposes  constraints to the parameter space of our model.
%
%some spectra even exceed the experimental constraints, so that we should be able to constrain the parameter space of this model.
\section{Experimental Discrimination between Dark Matter and Pulsar origin}
If indeed  the observed increase in the positron fraction is due to the pulsars  B0656 and Geminga, the fact that those two pulsars are located about 7\,degree apart on the sky in the outer Galaxy, we expect an anisotropy towards the galactic anticenter, as discussed in \cite{buesching08}.
 However, for DM we expect either an isotropic sky, or an anisotropy towards the galactic center region.

In the diffusion model, the anisotropy in the CR flux is given by \cite{1964ocr..book.....G}
$\delta={3\,k\left|\nabla N\right|}/({c\,N})$.
For instantaneous injection at time ${t_i}$ and distance ${r_i}$ for a diffusion coefficient as given by Eq.~\ref{efiff:k}, one arrives at 
\be
\delta=\frac{3}{2\,c}{ r_i}\,{ b_0}\, \left(\alpha-1  \right) {E}^{{\alpha}} \left( {E}^{\alpha -1}- \left( {E}/\left({\left(t-t_i\right)b_0E+1}\right)\right) ^{\alpha-1} \right) ^{-1},
\label{eq:ingo}
\ee 
\cite{buesching08} find an anisotropy of a few percent above 10\,GeV.
We  remark  that one expects for $E<10\,$GeV a solar cycle dependent anisotropy that is due to modulation of CR electrons and positrons.
\section{Summary}
The ATIC report of an excess just below 1\,TeV resulted in a wide interest in the
interpretation of the signal. This signal is also supported by the PAMELA detection
of an increase in the positron to electron ratio towards the energies where ATIC saw
their excess above the general cosmic ray lepton (electron and positron) spectral background.
Interestingly, this effect was not seen by the detector on the FERMI satellite\cite{ferminew}, nevertheless the latest FERMI results indicate a flattening of the CR electron spectrum above $\approx50\,GeV$, hinting at an additional leptonic component with a hard spectrum\cite{ferminew}.  
If we assume that the ATIC peak is real, then we are left with two competing explanations: 
(i) Kaluza-Klein type particle, or, (ii) a local source(s) of electrons and the best candidate
for the latter is a pulsar (with its PWN) since both the Kaluza-Klein and the pulsar
interpretations require both electrons and positrons, probably in equal numbers.
 
Whereas the Kaluza-Klein explanation required the introduction of a boost factor of 200 in the volume
production rate to explain the ATIC signal \cite{atic08}, the pulsar/PWN interpretation also involves a number of free parameters,
such as the birth period and interstellar diffusion coefficient between the source and earth.
%In this paper we have discussed a few scenarios which can fit the observed data and we are
%left with more questions than answers!
At the moment it seems as if we are left with too many free parameters and uncertainties
to make an unambiguous statement about the origin of the ATIC (or at least the PAMELA) signal.
More work needs to be done on DM candidate particles, as well as the multiwavelength
nature of pulsars and their wind nebulae. %New approaches may also be required to remove some of
%the uncertainties involved.   
%
%
%%%%%%%%%%%%%%%%%%%%%%%%%%%%%%%%%%%%%%%%%%%%%%%%%%%%%%%%%%%%%%%%%%%%%%%%%%%%
%
%
%\begin{figure}
%\begin{center}
%\psfig{file=procs-fig1.eps,width=4.5in}
%\end{center}
%\caption{The bifurcating response curves of system
%$\alpha=0.5, \beta=1.8; \delta=0.2, \gamma=0$: (a)
%$\mu=-1.3$; and (b) $\mu=0.3$.}
%\label{aba:fig1}
%\end{figure}
%

\end{document}